\documentstyle[12pt,epsf]{article}
\title{\bf Radiative leptonic decays of heavy mesons in heavy quark limit}
\author{Chien-Wen Hwang \footnote{Email:
t2732@nknucc.nknu.edu.tw} \\
{\small Department of Physics, National Kaohsiung Normal University,
Kaohsiung 824, Taiwan} \\
{\small Physics Division, National Center for Theoretical Sciences
(South), Tainan 701, Taiwan}}

\begin{document}
\maketitle
\begin{abstract}
We study the radiative leptonic decays of heavy mesons within the covariant light-front model. Using
this model, both the form factors $F_V$ and $F_A$ have the same form when the heavy quark limit is
taken. In addition, the relation between the form factor $F_V$ and the decay constant of heavy meson
$F_H$ is obtained. The hadronic parameter $\beta$ can be determine by the parameters appearing in the
wave function of heavy meson. We find that the value of $\beta$ is not only quite smaller than the
one in the non-relativistic case, but also insensitive to the value of light quark mass $m_q$. These
results mean that the relativistic effects are very important in this work. We also obtain the
branching ratio of $B\to l\nu_l \gamma$ is about $(1.40-1.67)\times 10^{-6}$, in agreement with the
general estimates in the literature.
\end{abstract}
\newpage

\section{Introduction}
The understanding of the strong and weak interactions of a heavy quark system is an important topic,
and the purely leptonic decays $B\to l\nu_l$ seems to be the useful tools for this purpose. In
particular, these processes are very simple in that no hadrons and photons appear in the final
states. However, the rates of these purely leptonic decays are helicity suppressed with the factor of
$M_l^2/M_B^2$ for $l=e$ and $\mu$ (the $\tau$ channel, in spite of no suppression, is hard to observe
the decay because of the low efficiency). Therefore it is natural to extend the purely leptonic $B$
decay searches to the corresponding radiative modes $B\to l\nu_l \gamma$. These radiative leptonic
decays receive two types of contributions: inner bremsstrahlung (IB) and structure-dependent (SD)
\cite{GW,Wyler}. As is known, the IB contributions are still helicity suppressed, while the SD ones
are reduced by the fine structure constant $\alpha$ but they are not suppressed by the lepton mass.
Accordingly, the radiative leptonic $B$ decay rates could have an enhancement with respect to the
purely leptonic ones, and would offer useful information about the Cabibbo-Kobayashi-Maskawa matrix
element $|V_{ub}|$ and the decay constant $f_B$ \cite{CFN}. Recently, there has been a great deal of
theoretical attention \cite{Wyler,CFN,KSW,EHM,AES,geng,Yan,CL} to these radiative leptonic $B$
decays. Experimentally, the current upper limits for these modes are $Br(B^+\to e^+\nu_e\gamma)<
2.2\times 10^{-5}$ and $Br(B^+\to \mu^+\nu_\mu \gamma)< 2.3\times 10^{-5}$ at the $90\%$ confidence
level \cite{Belle}. With the better statistic expected from the $B$ factories, the observations of
these decays could become soon feasible.

The hadronic matrix elements responsible for the above decays can be calculated in various quark
models. However, the relativistic effects must be considered seriously in calculations as the recoil
momentum is large. This problem is taken into account by the light-front quark model (LFQM)
\cite{lfqm} which has been considered as one of the best effective relativistic quark models in the
description of the exclusive heavy hadron decays \cite{HCC}. Its simple expression, relativistic
structure and predictive power have made wide applications of the LFQM in exploring and predicting
the intrinsic heavy hadron dynamics. However, almost all the previous investigations have not
covariantly extracted the form factors from the relevant matrix elements and paid enough attention to
the consistency with heavy quark symmetry (HQS) and heavy quark effective theory (HQET). The
covariant light-front model \cite{HCCZ} has resolved the above-mentioned shortcomings in the LFQM and
has improved the current understanding of the QCD analysis of heavy hadrons. This model consists of a
heavy meson bound state in the heavy quark limit (namely $m_Q\to \infty$), which is fully consistent
with HQS, plus a reliable approach from this bound state to systematically calculate the $1/m_Q$
corrections within HQET in terms of the $1/m_Q$ expansion of the fundamental QCD theory. In this
paper, we will use the covariant light-front model to investigate the radiative leptonic $B$ decays
in the heavy quark limit.

The paper is organized as follows. In Sec. 2 a general construction of covariant light-front bound
states is provided, the diagrammatic rules within this model are also listed. In Sec. 3 we evaluate
the decay constant of heavy meson $F_H$ and the form factors of the radiative leptonic heavy meson
decay $F_{V,A}$ in a complete covariant way. In Sec. 4 the relation between $F_H$ and $F_V$ is
obtained. A few numerical calculations made with the help of the Gaussian-type wave function are
presented. Finally, a summary is given in Sec. 5.

\section{Covariant light-front model}
The light-front bound states of heavy meson that are written in a form of exhibiting explicitly the
boost covariance have been shown in the literature \cite{HCC}. In this paper, we focus on the bound
states of heavy mesons in the heavy quark limit:
\begin{eqnarray}  \label{hqslfb}
  |H(v;S,S_z)\rangle &=& \int [d^3k][d^3p_q] 2(2\pi)^3v^+ \delta^3(
    \overline{\Lambda}v-k-p_q) \nonumber \\
    &\times& \sum_{\lambda_Q,\lambda_q}
    R^{SS_z} (X, \kappa_{\bot}, \lambda_Q, \lambda_q) \Phi^{SS_z} (X,\kappa^2_{\bot})
    b_v^\dagger(k, \lambda_Q) d_q^\dagger (p_q, \lambda_q)|0\rangle,
\end{eqnarray}
where $v^\mu~(v^2=1)$ is the velocity of the heavy meson, $k=p_Q-m_Q v$ is the residual momentum of
heavy quark, $p_q$ is the momentum of light antiquark,
\begin{equation}
[d^3k] ={ dk^+ d^2 k_\bot \over{ 2(2\pi)^3 v^+}},~~[d^3p_q] ={dp_q^+ d^2 p_{q\bot} \over{ 2(2\pi)^3
p_q^+}},
\end{equation}
and $\overline{\Lambda}=M_H-m_Q$ is the residual center mass of heavy mesons. The relative momentum
$X$ was first introduced in Ref. \cite{CZL} as the product of longitudinal momentum fraction $x$ of
the valence antiquark and the mass of heavy meson $M_H$, namely $X=x M_H$. The relative transverse
and longitudinal momenta, $\kappa_\bot$ and $\kappa_z$, are obtained by
\begin{equation}
\kappa_\bot=p_{q\bot}-X v_\bot,~~\kappa_z={X\over{2}}-{m_q^2+p_{q\bot}^2\over{2 X}}.\label{kzH}
\end{equation}
In Eq. (\ref{hqslfb}), $\lambda_Q$ and $\lambda_q$ are helicities of heavy quark and light antiquark,
respectively. In phenomenological calculations, one usually ignores the dynamical dependence of the
light-front spin so that the function $R^{S,S_z}$ can be approximately expressed by taking the
covariant form for the so-called Melosh matrix \cite{melosh} in the heavy quark limit,
\begin{equation}  \label{spin}
  R^{SS_z}(X, \kappa_{\bot}, \lambda_Q, \lambda_q)
    = \cases{ {1\over 2} \sqrt{ 1
        \over v\cdot p_q + m_q} ~ \overline u_{\lambda_Q}(v)
        (i\gamma^5) v_{\lambda_q}(p_q) ~~~~ &{\rm for}~~ S=0, \cr
        {1\over 2} \sqrt{ 1 \over v\cdot p_q
        + m_q}~ \overline u_{\lambda_Q}(v)
        (-\! \! \not \! \epsilon) v_{\lambda_q}(p_q)
        ~~~~ & {\rm for}~~ S=1, \cr}
\end{equation}
where $u_{\lambda_Q}(v)$ and $v_{\lambda_q}(p_q)$ are spinors for the heavy quark and light
antiquark,
\begin{equation}
    \sum_\lambda u_{\lambda}(v)\overline{u}_{\lambda}(v)
        = {{\not \! v}+1},~~\sum_\lambda v_{\lambda}(p_q)\overline{v}_{\lambda}(p_q)
        = {{\not \! p_q}-m_q}.
\end{equation}
The operators $b_v^\dagger(k,\lambda_Q)$ and $d^\dagger_q (p_q,\lambda_q$) create a heavy quark and a
light antiquark with
\begin{eqnarray}
    \{ b_v (k,\lambda_Q), ~ b_{v'}^\dagger (k',\lambda'_Q) \}
        &=&2 (2\pi)^3 v^+ \delta_{vv'}\delta^3(k-k')
        \delta_{\lambda_Q \lambda'_Q},\nonumber \\
    \{ d_q (p_q,\lambda_q), ~ d_q^\dagger (p'_q,\lambda'_q) \}
        &=&2 (2\pi)^3 p_q^+ \delta^3(p_q-p'_q)
        \delta_{\lambda_q \lambda'_q}.
\end{eqnarray}
The normalization of the heavy meson bound states in the heavy quark limit is then given by
\begin{equation}  \label{nmc2}
    \langle H(v',S',S'_z) |H(v,S,S_z)\rangle = 2(2\pi)^3 v^+
        \delta^3(\overline{\Lambda}v'-\overline{\Lambda}v)
        \delta_{SS'} \delta_{S_zS'_z},
\end{equation}
which leads to two things: first, the heavy meson bound state $|H(v;S,S_z)\rangle$ in this model
rescales the one $|\tilde {H}(P;S,S_z)\rangle$ in the LFQM by $|\tilde
{H}\rangle=\sqrt{M_H}|H\rangle$ and, second, the space part $\Phi^{SS_Z}(X,\kappa_\bot^2)$ (called
the light-front wave function) in Eq.(\ref{hqslfb}) has the following wave function normalization
condition:
\begin{equation} \label{nwf}
    \int {dX d^2\kappa_\bot \over 2(2\pi)^3 X}
        |\Phi^{SS_z} (X, \kappa^2_\bot)|^2 = 1.
\end{equation}
In principle, the heavy quark dynamics is completely described by HQET, which is given by the $1/m_Q$
expansion of the heavy quark QCD Lagrangian
\begin{equation}
    {\cal L} = \overline{Q} (i \not \! \! D - m_Q) Q \nonumber
    = \sum_{n=0}^\infty \Bigg({1 \over 2m_Q}
        \Bigg)^n {\cal L}_n.
        \label{hqcdl}
\end{equation}
Therefore, $|H(v;S,S_z)\rangle$ and $\Phi^{SS_z} (X,\kappa^2_{\bot})$ are then determined by the
leading Lagrangian ${\cal L}_0$. The authors of Ref. \cite{HCCZ} have shown from the light-front
bound state equation that $\Phi^{SS_z} (X,\kappa^2_{\bot})$ must be degenerate for $S=0$ and $S=1$.
As a result, we can simply write
\begin{equation} \label{msiwf}
    \Phi^{SS_z}(X,\kappa^2_\bot) = \Phi(X,\kappa^2_\bot)
\end{equation}
in the heavy quark limit. Eq.~(\ref{hqslfb}) together with
Eqs.~(\ref{spin}) and (\ref{msiwf}) is then the heavy meson
light-front bound states in the heavy quark limit that obey HQS.
Furthermore, Eq.~(\ref{hqslfb}) can be rewritten in a fully
covariant form if $\Phi(X,\kappa^2_\bot)$ is a function of $v \cdot
p_q$:
\begin{equation} \label{cwf}
    \Phi (X, \kappa^2_\bot) \longrightarrow \Phi (v \cdot p_q) \, ,
\end{equation}
where the antiquark $q$ in bound states is on-mass-shell, $p_q^- =
 (p_{q\bot}^2 + m_q^2)/p_q^+$. Hence,
\begin{equation}\label{vponshell}
    v \cdot p_q = {1\over 2X} \Big( \kappa_\bot^2 + m_q^2
        + X^2 \Big) \, .
\end{equation}
As to the normalization condition of $\Phi (v \cdot p_q)$, Eq.~(\ref{nwf}) can also be rewritten in a
covariant form:
\begin{equation} \label{cn}
    \int {d^4 p_q \over (2\pi)^4} (2\pi) \delta (p_q^2 - m_q^2)
        | \Phi ( v\cdot p_q) |^2 = 1.
\end{equation}
The left-hand side of Eq. (\ref{cn}) can be easily obtained in a diagrammatic way as shown in Fig.~1:
\begin{eqnarray}
{\rm Fig.~1} &=& \int {d^4 p_q \over (2\pi)^4} (2\pi) \delta( p_q^2
    - m_q^2) |\Phi (v \cdot p_q)|^2 {{\rm Tr} \Big[ \Gamma_H (1 + \not \! v) \Gamma_H
        ( \not \! p_q - m_q) \Big] \over 4(v \cdot p_q
    + m_q)}\nonumber \\
        &=& {\rm Eq.~(\ref{cn})} \, .
\end{eqnarray}
In general, the on-shell Feynman rules within this model is given as follows \cite{HCCZ}:

(i) The heavy meson bound state in the heavy quark limit gives a vertex as follows:
\begin{equation}
\begin{picture}(65,30)(0,38)
\put(20,41){\line(1,0){20}} \put(20,39){\line(1,0){20}}
\put(40,40){\circle*{6} }
\end{picture}
    : ~~ {1\over 2} \sqrt{1 \over v \cdot p_q + m_q}
        ~\Phi(v \cdot p_q) \Gamma_H \, .
\end{equation}

(ii) The internal line attached to the bound state gives an on-shell propagator:
\begin{eqnarray}
\begin{picture}(65,30)(0,38)
\put(0,40.5){\line(1,0){40}} \put(19,40){\vector(1,0){2}} \put(0,39){\line(1,0){40}} \put(19,28){$k$}
\end{picture}
    &:& ~~ {1 + \not \! v} ~~ ({\rm for~heavy~quarks}), ~~~ \\
\begin{picture}(65,30)(0,38)
\put(0,40){\line(1,0){40}} \put(21,40){\vector(-1,0){2}}
\put(17.5,30){$p_q$}
\end{picture}
    &:& ~~ { \not \! p_q - m_q} ~~ (\rm for ~light~antiquarks).
\end{eqnarray}

(iii) For the internal antiquark line attached to the bound state, sum over helicity and integrate
the internal momentum using
\begin{equation}
     \int {d^4 p_q \over (2\pi)^4 } (2\pi) \delta
        (p_q^2 -m_q^2).
\end{equation}

(iv) For all other lines and vertices that do not attach to the bound states, the diagrammatic rules
are the same as the Feynman rules in the conventional field theory.

\section{Decay constants and form factors}
Now, we shall present the evaluations of the decay constants and the form factors for heavy mesons
within the covariant light-front model. First, the decay constants of pseudoscalar and vector mesons
are defined by $\langle 0 | \overline{q} \gamma^\mu \gamma_5 Q |\tilde {H} \rangle = i f_H p^\mu$ and
$\langle 0 | \overline{q}\gamma^\mu Q |\tilde {H}^* \rangle = f_{H^*} M_{H^*} \epsilon^\mu$, where
$q$ and $Q$ the light and heavy quark field operators, respectively. In the heavy quark limit, the
decay constants have the expressions
\begin{equation}
    \langle 0 | \overline{q} \gamma^\mu \gamma_5 h_v | H(v)
        \rangle = i F_H v^\mu~~,
        ~~~ \langle 0 | \overline{q}\gamma^\mu h_v |
        H^*(v,\epsilon) \rangle = F_{H^*} \epsilon^\mu \, ,
\end{equation}
so that
\begin{equation}
    F_H = f_H \sqrt{M_H}~~,~~~ F_{H^*} = f_{H^*} \sqrt{M_{H^*}} \, .
\end{equation}
Using the above bound states and Feynman rules, it is very simply to
evaluate the relevant matrix elements (diagrammatically shown in
Fig.~2):
\begin{eqnarray}
\langle 0 | \overline{q} \gamma^\mu \gamma_5 h_v |H (v) \rangle
    &=& - i {\rm Tr}\Big\{\gamma^\mu \gamma_5 {\not \! v + 1
        \over 2} \gamma_5 {\cal M}_1 \Big\} \, , \\
\langle 0 | \overline{q} \gamma^\mu h_v |H^*(v, \epsilon) \rangle
        &=& {\rm Tr}\Big\{\gamma^\mu {\not \! v + 1
        \over 2} \not \! \epsilon {\cal M}_1 \Big\} \, ,
\end{eqnarray}
where
\begin{equation}  \label{covint1}
    {\cal M}_1 = \sqrt{N_c} \int {d^4 p_q \over (2\pi)^4} (2\pi)
        \delta(p_q^2- m_q^2) {\Phi(v \cdot p_q) \over
        \sqrt{v \cdot p_q +m_q}} ( m_q - \not \! p_q)
        = A_1 + B_1 \! \not \! v \, ,
\end{equation}
and
\begin{eqnarray}
    A_1 &=& \sqrt{N_c} \int {d^4 p_q \over (2\pi)^4} (2\pi)
        \delta(p_q^2- m_q^2) {\Phi( v \cdot p_q) \over
        \sqrt{v \cdot p_q +m_q}}~ m_q ~~,  \label{A1}\\
    B_1 &=& - \sqrt{N_c} \int {d^4 p_q \over (2\pi)^4}(2\pi)
        \delta(p_q^2- m_q^2) {\Phi(v \cdot p_q) \over
        \sqrt{v \cdot p_q +m_q}} ~v \cdot p_q \, .\label{B1}
\end{eqnarray}
Here $N_c=3$ is the number of colors. Thus, it is easily found:
\begin{eqnarray}
F_H &=&  2 (A_1 - B_1) \nonumber \\
&=& 2\sqrt{N_c}\int {d^4 p_q \over (2\pi)^4}
        (2\pi) \delta(p_q^2- m_q^2) {\Phi( v \cdot p_q) \over
        \sqrt{v \cdot p_q +m_q}} (v \cdot p_q +m_q)=F_{H^*}, \label{decc}
\end{eqnarray}
as expected from HQS. Using Eq. (\ref{vponshell}), the integral in Eq. (\ref{decc}) gives
\begin{equation}
F_H=2\sqrt{2 N_c} \int {dXd^2 \kappa_\bot \over 2(2\pi)^3 \sqrt{X}}
        ~{\Phi(X,\kappa^2_\bot)\over{\sqrt{\kappa_\bot^2 +
        (m_q+X)^2}}}\Bigg({X\over{2}}+{m_q^2+\kappa_\bot^2\over{2 X}}+m_q\Bigg). \label{decc1}
\end{equation}

Next, the form factors for the radiative leptonic decays, which come from vector and axial vector
currents are defined by \cite{EHM}
\begin{eqnarray}
\langle\gamma(p_\gamma,\epsilon) |\bar q\gamma_\mu Q| \tilde{H}(P)\rangle &=&e
{f_V(q^2)\over{M_H}}\varepsilon_{\mu\nu\alpha\beta}\epsilon^{*\nu}P^{\alpha}p_\gamma^{\beta}, \label{fv}\\
\langle\gamma(p_\gamma,\epsilon) |\bar q\gamma_\mu \gamma_5 Q| \tilde{H}(P)\rangle &=&
ie{f_A(q^2)\over{M_H}}[\epsilon_\mu^* (P\cdot p_\gamma) - (p_\gamma)_\mu (\epsilon^*\cdot P)],
\label{fa}
\end{eqnarray}
respectively, where $q=P-p_\gamma$. If one ignores the lepton mass (namely ignores the IB
contributions), the leading contributions to $f_V(q^2)$ come from pole diagrams with one vector
intermediate state ($J^P=1^-$), and those to $f_A(q^2)$ from two axial vector states ($J^P=1^+$)
\cite{Wyler}. In the non-relativistic quark model (NRQM), the dominant contribution comes from the
former state,
\begin{equation}
f_V(q^2)={q_q\over{2(E_\gamma+\triangle)}}~\beta f_{H^*} M_{H^*},
\end{equation}
where $q_q$ is the charge of light quark in units of $e$, $\triangle=M_{H^*}-M_H$, and the hadronic
parameter $\beta\simeq 3$ Gev$^{-1}$ \cite{ABJLM}. The contributions of the latter states are also
proportional to $f_A^i$, the axial vector meson decay constants. But these are zero because the wave
function at the origin for an orbitally excited state vanishes.

In the heavy quark limit, the above form factors have the expressions
\begin{eqnarray}
\langle\gamma(p_\gamma,\epsilon) |\bar q\gamma_\mu h_v| H(v)\rangle
&=&e
F_V(E_\gamma)\varepsilon_{\mu\nu\alpha\beta}\epsilon^{*\nu}v^{\alpha}p_\gamma^{\beta},\\
\langle\gamma(p_\gamma,\epsilon) |\bar q\gamma_\mu \gamma_5 h_v|
H(v)\rangle &=& ie F_A(E_\gamma)[\epsilon_\mu^* (v\cdot p_\gamma) -
(p_\gamma)_\mu (\epsilon^*\cdot v)],
\end{eqnarray}
where $E_\gamma=v\cdot p_\gamma$. So that
\begin{equation}
F_i={f_i\over{\sqrt{M_H}}}\,\,\,({\rm for}~~~i=V, A).
\end{equation}
The contributions to these form factors coming from the coupling of the photon to the heavy and light
quark can be diagrammatically shown in Fig.~3 (a) and (b), respectively. Using the above bound states
and Feynman rules, the former contributions to the form factor, for example, $F_V$ are given by
\begin{eqnarray} \label{fvh}
\Gamma_\mu^{(Q)}=q_Q{ie\over{2m_Q v\cdot p_\gamma}}{\rm Tr}\Big\{\gamma_\mu (m_Q {\not v}-{\not \!
p_\gamma}+m_Q){\not \epsilon}{ \not \! v + 1 \over 2}\gamma_5 {\cal M}_1\Big\},
\end{eqnarray}
where $q_Q$ is the charge of heavy quark in units of $e$ and ${\cal M}_1$ was obtained in Eq.
(\ref{covint1}). Performing the trace and comparing with Eq. (\ref{decc}), the heavy quark
contribution to the form factor $F_V$ is obtained by
\begin{equation}
F_V^{(Q)}(E_\gamma)={q_Q\over{2E_\gamma}}{2(A_1-B_1)\over{m_Q}}={q_Q\over{2E_\gamma}}{F_H\over{m_Q}},
\end{equation}
as obtained from perturbative QCD \cite{Yan}. In the case of the
light quark contribution, it can be written as
\begin{eqnarray}\label{fvq}
\Gamma_\mu^{(q)}=i q_q {\rm Tr}\Big\{\gamma_\mu { \not \! v + 1 \over 2} \gamma_5 {\cal M}_2 {\not \!
p_\gamma}{\not \epsilon}\Big\},
\end{eqnarray}
where
\begin{equation} \label{M2}
{\cal M}_2 = \sqrt{N_c} \int {d^4 p_q \over (2\pi)^4} (2\pi)
        \delta(p_q^2- m_q^2) {\Phi(v \cdot p_q) \over
        \sqrt{v \cdot p_q +m_q}} {( m_q - \not \! p_q)\over{2p_\gamma\cdot p_q}}
        = A_2 + B_2 \! \not \! v + C_2 \! \not \! p_\gamma~,
\end{equation}
and
\begin{eqnarray}
A_2 &=& \sqrt{N_c} \int {d^4 p_q \over (2\pi)^4} (2\pi)
        \delta(p_q^2- m_q^2) {\Phi( v \cdot p_q) \over
        \sqrt{v \cdot p_q +m_q}}~ {m_q \over{2p_\gamma\cdot p_q}}
        ~,  \label{A2} \\
    B_2 &=& \sqrt{N_c} \int {d^4 p_q \over (2\pi)^4}(2\pi)
        \delta(p_q^2- m_q^2) {\Phi(v \cdot p_q) \over
        \sqrt{v \cdot p_q +m_q}} ~{1\over{2 v \cdot p_\gamma}}~,
        \label{B2} \\
    C_2 &=& \sqrt{N_c} \int {d^4 p_q \over (2\pi)^4}(2\pi)
        \delta(p_q^2- m_q^2) {\Phi(v \cdot p_q) \over
        \sqrt{v \cdot p_q +m_q}} ~{1\over{2 v \cdot p_\gamma}}
        {v\cdot p_q v\cdot p_\gamma-p_q\cdot p_\gamma\over{v \cdot p_\gamma
        p_q\cdot p_\gamma }}~.\label{B2}
\end{eqnarray}
It is easily found
\begin{eqnarray}
F_V^{(q)} &=& 2 q_q (A_2 - B_2) \nonumber\\
&=& {q_q \over{2 E_\gamma}}\sqrt{N_c}\int {d^4 p_q \over (2\pi)^4}
        (2\pi) \delta(p_q^2- m_q^2) {\Phi( v \cdot p_q) \over
        \sqrt{v \cdot p_q +m_q}} 2\Bigg(1+m_q {v \cdot p_\gamma\over{p_q\cdot p_\gamma}}\Bigg). \label{FVq}
\end{eqnarray}
Using the light-front relative momentum, the integral Eq.
(\ref{FVq}) gives
\begin{equation}\label{FV3}
F_V^{(q)}={q_q \over{2 E_\gamma}} 2\sqrt{2 N_c} \int {dXd^2
\kappa_\bot \over 2(2\pi)^3 \sqrt{X}}
        ~{\Phi(X,\kappa^2_\bot)\over{\sqrt{\kappa_\bot^2 +
        (m_q+X)^2 }}}\Bigg(1+{m_q\over{m_q^2+\kappa_\bot^2\over{X}}}\Bigg).
\end{equation}
With this way, we also calculate the form factors $F_A^{(q,Q)}$ which come from the coupling to the
axial vector current and find that $F_A^{(q,Q)}(E_\gamma)=F_V^{(q,Q)}(E_\gamma)$. These results are
consistent with those in Ref. \cite{Yan}, but contrary to those in NRQM \cite{Wyler}. In addition, it
has been emphasized in the literature that various hadronic form factors calculated in the LFQM
should be extracted only from the plus component of the corresponding currents. Also, the LFQM
calculations for decay processes are restricted only with a specific Lorentz frame (namely zero
momentum transfer). Now we see that the covariant light-front model removes the above restrictions,
straightforwardly extracts the hadronic form factors $F_{V,A}$, and obtains the result $F_H=F_{H^*}$
which is consistent with HQS.

\section{Numerical calculations and discussions}
At first glance, the decay constant $F_H$ in Eq. (\ref{decc1}) doesn't seems to connect with
$F_V^{(q)}$ in Eq. (\ref{FV3}). However, since the wave function $\Phi(X,\kappa_\bot^2)$ is the
function of $v\cdot p_q$ (see Eq. (\ref{cwf})) for a fully covariant bound state, and $v\cdot
p_q=E_q=\sqrt{m_q^2+\kappa_\bot^2+\kappa_z^2}$, thus $\Phi(X,\kappa_\bot^2)$ is even in $\kappa_z$.
It follows that
\begin{eqnarray}
\int {dXd^2 \kappa_\bot \over 2(2\pi)^3 \sqrt{X}}
        ~{\Phi(X,\kappa^2_\bot)\over{\sqrt{\kappa_\bot^2 +
(m_q+X)^2 }}}~\kappa_z=0. \label{odd}
\end{eqnarray}
From Eqs. (\ref{kzH}) and (\ref{odd}), $F_H$ and $F_V^{(q)}$ are easily rewritten in the simpler
forms:
\begin{eqnarray}
F_H &=& 2\sqrt{2 N_c} \int {dXd^2 \kappa_\bot \over 2(2\pi)^3
\sqrt{X}}
        ~{\Phi(X,\kappa^2_\bot)\over{\sqrt{\kappa_\bot^2 +
        (m_q+X)^2}}}(X+m_q),  \\
F_V^{(q)}&=&{q_q \over{2 E_\gamma m_q}} 2\sqrt{2 N_c} \int {dXd^2
\kappa_\bot \over 2(2\pi)^3 \sqrt{X}}
        ~{\Phi(X,\kappa^2_\bot)\over{\sqrt{\kappa_\bot^2 +
        (m_q+X)^2 }}}(X+m_q)\Big({m_q\over{X}}\Big). \label{FVFH}
\end{eqnarray}
It is known that $X v^+=p_q^+=E_q+\kappa_z$, and therefore the ratio ($m_q/X$) in Eq. (\ref{FVFH})
will be equal to unity when the antiquark is at rest. Thus, in the non-relativistic case, Eq.
(\ref{FVFH}) can be reduced to
\begin{equation}
F_V^{(q)}={q_q \over{2 E_\gamma }}{F_H\over{m_q}}. \label{NRQM}
\end{equation}
From Eq. (\ref{NRQM}), one can extract the hadronic parameter $\beta_{NR}=1/m_q$ which is consistent
with the result of the NRQM \cite{Wyler,ABJLM}. In the relativistic case, we can evaluate Eq.
(\ref{FVFH}) directly to extract $\beta_R$. There are serval popular phenomenological wave function
that have been employed to describe various hadronic structure in the literature. We choose the
Gaussian-type wave function $\Phi_G (v\cdot p_q)$ \cite{HCCZ} which have been widely used in the
study of heavy mesons:
\begin{equation}
\Phi_G(v\cdot p_q)=4 \Big({\pi\over \omega^2}\Big)^{3/4}
        \sqrt{ v \cdot p_q} \exp\Bigg\{-{1\over 2\omega^2}
        \Big[ (v \cdot p_q)^2 - m_q^2 \Big]\Bigg\} \
\end{equation}
The parameters appearing in this wave function are $m_q$ and $\omega$. From the NRQM to the LFQM, the
value of light quark mass varies from $0.33$ to $0.23$ GeV. Thus we take several different values of
$m_q$ and $f_B$ to evaluate the parameter $\beta_R$. These results are listed in Table I. We find
that, in general, the values of $\beta_R$ are not only quite smaller than the ones of $\beta_{NR}$,
but also insensitive to the values of $m_q$. These results mean that the typical values of $X$ in the
integrations are much larger than $m_q$, that is to say, the relativistic effects are very important
in these evaluations. \vskip 0.2cm {\small Table I. Parameters $m_q$ and $\omega$ fitted to the decay
constant $f_B$. Also shown are the results of hadronic parameter $\beta_R$ and $\beta_{NR}$.}
\begin{center}
\begin{tabular}{c | c c c | c c c}
\hline\hline $f_B$ (GeV) & & 0.180 & & & 0.190 & \\
$m_q$ (GeV) & 0.33 & 0.28 & 0.23& 0.33 & 0.28 & 0.23 \\
$\omega$ (GeV) & 0.480 & 0.486 & 0.493& 0.499 & 0.505& 0.513\\
\hline $\beta_R$ (GeV$^{-1}$) &1.58 & 1.66 & 1.73 & 1.55 & 1.61 & 1.67\\
$\beta_{NR}$ (GeV$^{-1}$) & 3.03 & 3.57 & 4.35 & 3.03 & 3.57 & 4.35\\\hline \hline \end{tabular}
\end{center}
This hadronic parameter $\beta_R$ can be used to calculate the branching ratio of the $B$ meson
radiative decay. The decay rate for $B\to l\nu_l \gamma$ differential in the photon energy is given
by
\begin{equation}
{d\Gamma\over{dE_\gamma}}={\alpha G_F^2|V_{ub}|^2
M_B^4\over{48\pi^2}}[f_V^2(E_\gamma)+f_A^2(E_\gamma)]y^3(1-y),
\label{Br}
\end{equation}
where $y\equiv 2E_\gamma /M_B$. In the heavy quark limit, there is
only one contribution $F_{V,A}^{(q)}$ to $F_{V,A}$, thus $f_{V,A}$
in Eqs. (\ref{fv}) and ({\ref{fa}}) can be obtained by
\begin{equation}
f_{V,A}={q_q\over{2E_\gamma}}\beta_R f_B M_B.
\end{equation}
Integrating the photon energy in Eq. (\ref{Br}), we can obtain the decay rate
\begin{equation}
\Gamma(B\to l\nu_l \gamma)={\alpha G_F^2|V_{ub}|^2
M_B^5\over{288\pi^2}}~q_q^2 \beta_R^2 f_B^2.
\end{equation}
Taking $f_B=0.18$ GeV, $|V_{ub}|^2=3.33\times 10^{-3}$ \cite{PDG04}, and $\tau_{B^+}=1.67\times
10^{-12}$ sec \cite{PDG04}, we obtain $Br(B\to l\nu_l \gamma)=(1.40-1.67)\times 10^{-6}$. This result
agrees well with that in Refs. \cite{EHM} and \cite{geng} where the light cone QCD sum rules and the
LFQM were used in their calculations, respectively. However, this result is about a factor of 2
smaller and larger than that in Refs. \cite{AES} and \cite{CFN}, respectively. The parameter
$\beta_R$ also relates to a ratio $Br(B\to l\nu_l \gamma)/Br(B\to \mu \nu_\mu)$. The pure leptonic
decay rate is given by
\begin{equation}
\Gamma(B\to \mu \nu_\mu)={G_F^2|V_{ub}|^2 M_B^3\over{8\pi^2}}~ f_B^2
\Bigg({M_\mu^2\over{M_B^2}}\Bigg)\Bigg(1-{M_\mu^2\over{M_B^2}}\Bigg).
\end{equation}
Taking the relevant values, we obtain the ratio
\begin{equation}
{Br(B\to l\nu_l \gamma)\over{Br(B\to \mu \nu_\mu)}}\simeq 2.0
\beta_R^2 \simeq 5.0-6.0,
\end{equation}
which is within the range of $1-30$ as expected in Ref.
\cite{Wyler}.

\section{Summary and perspective}
In this work we have calculated the decay constants $F_{H,H^*}$ for heavy mesons and the form factors
$F_{V,A}(E_\gamma)$ for the radiative leptonic decays $H\to l\nu\gamma$ within the covariant
light-front approach. In accordance with HQS and the results of Ref. \cite{Yan}, $F_H=F_{H^*}$ and
$F_V(E_\gamma)=F_A(E_\gamma)$ to the tree level, respectively. In addition, the form factor
$F_V(E_\gamma)$ can be related to the decay constant $F_{H^*}$ by considering an $1^-$ intermediate
state contribution. The relevant hadronic parameter $\beta_R$, in contrast to $\beta_{NR}=1/m_q$ in
the NRQM, has been determined by the parameters $m_q$ and $\omega$ in this covariant model. The
comparisons between $\beta_R$ and $\beta_{NR}$ was listed quantitatively in Table I. We conclude that
the relativistic effects are quite obvious in these modes. We have also obtained $Br(B\to l\nu_l
\gamma)=(1.40-1.67)\times 10^{-6}$ and $Br(B\to l\nu_l \gamma)/Br(B\to \mu \nu_\mu)=5.0-6.0$, in
agreement with the general estimates in the literature.

The covariant model removes some restrictions often occurred in the usual LFQM calculation, and, as
we have shown in this paper, allows one to perform some extremely simple evaluation of various heavy
meson properties. Further applications to other properties of heavy mesons and physical processes
will be presented in subsequent papers.

\section*{Acknowledgement}
I am grateful to Chuang-Hung Chen for valuable discussions. I also
wish to thank the National Center for Theoretical Sciences (South)
for its hospitality during my summer visit where this work started.
This work was supported in part by the National Science Council of
R.O.C. under Grant No. NSC93-2112-M-017-001.

\newpage
\begin{figure}
\vskip 4.cm \includegraphics{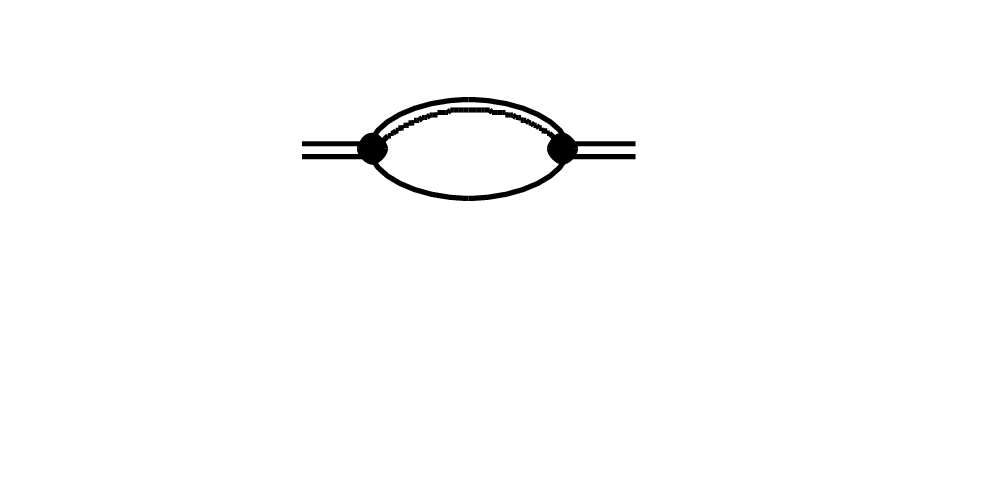} \caption{A
diagrammatic form for the heavy meson state normalization.}
\end{figure}

\begin{figure}
\vskip 4.cm \includegraphics{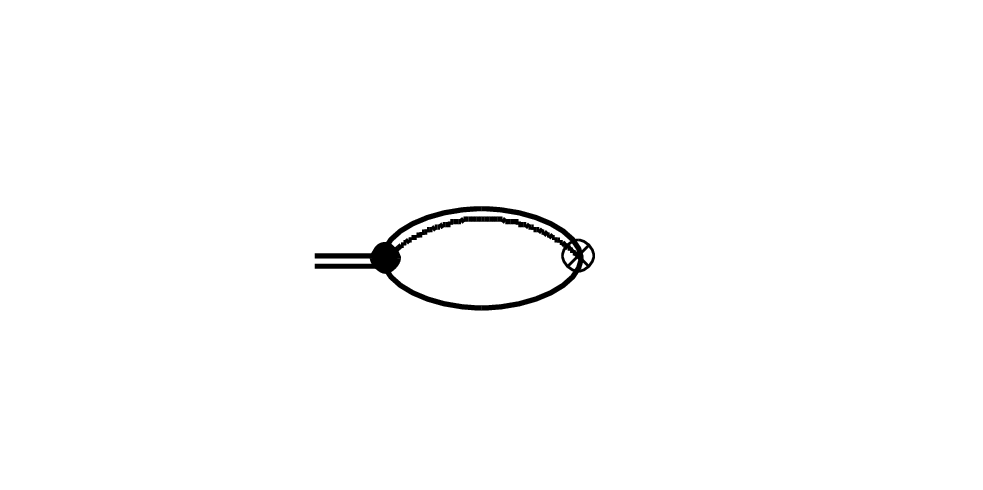} \caption{The
diagram for heavy meson decays.}
\end{figure}

\begin{figure}
\vskip 4.cm \includegraphics{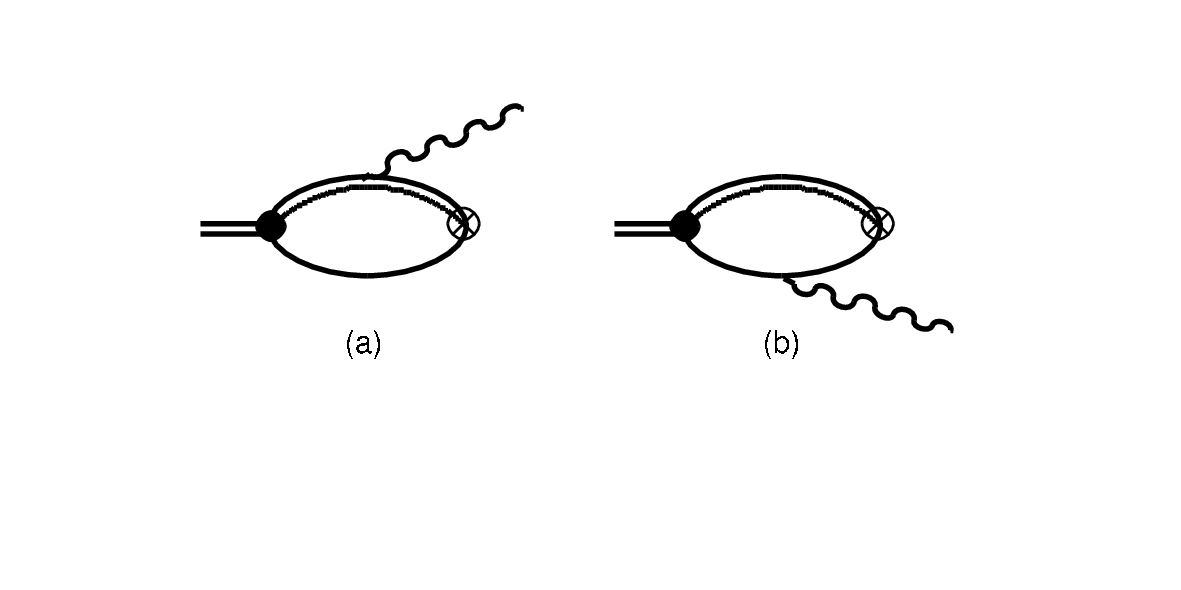} \caption{The
diagrams for the radiative leptonic decays of heavy meson.}
\end{figure}

\end{document}